\documentclass[useAMS,usenatbib]{mn2e}

\usepackage{latexsym,graphicx,natbib}

\usepackage{color}

\newcommand\kms{{\rm\,km\,s^{-1}}}
\newcommand\msun{\rm\,M_\odot}

\newcommand\lsun{\rm\,L_\odot}
\newcommand\hii{H\,{\sc ii} \,}

\def\apgt{\ {\raise-.5ex\hbox{$\buildrel>\over\sim$}}\ }
\def\aplt{\ {\raise-.5ex\hbox{$\buildrel<\over\sim$}}\ }

\title[New Galactic bona fide luminous blue variable]{Discovery of a new Galactic bona fide luminous blue variable with {\it Spitzer}\footnotemark[0]\thanks{Based on observations
made with the Southern African Large Telescope.}}
\author[V.V.Gvaramadze et al.]
       {V. V.~Gvaramadze,$^{1,2,3}$\thanks{E-mail: vgvaram@mx.iki.rssi.ru (VVG)},
        A. Y.~Kniazev,$^{4,5,1}$, L. N.~Berdnikov,$^{1,6,3}$ N.~Langer,$^{7}$
        \newauthor
        E. K.~Grebel$^{8}$ and J. M.~Bestenlehner$^{7}$\\
        $^{1}$Sternberg Astronomical Institute, Lomonosov Moscow State University, Universitetskij Pr. 13, Moscow 119992, Russia\\
        $^{2}$Space Research Institute, Russian Academy of Sciences, Profsoyuznaya 84/32, 117997 Moscow, Russia \\
        $^{3}$Isaac Newton Institute of Chile, Moscow Branch, Universitetskij Pr. 13, Moscow 119992, Russia \\
        $^{4}$South African Astronomical Observatory, PO Box 9, 7935 Observatory, Cape Town, South Africa \\
        $^{5}$Southern African Large Telescope Foundation, PO Box 9, 7935 Observatory, Cape Town, South Africa \\
        $^{6}$Astronomy and Astrophysics Research division,  Entoto Observatory and Research Center, P.O.Box 8412, Addis Ababa, Ethiopia \\
        $^{7}$Argelander-Institut f\"ur Astronomie der Universit\"at Bonn, Auf dem H\"ugel 71, 53121, Bonn, Germany \\
        $^{8}$Astronomisches Rechen-Institut, Zentrum f\"{u}r Astronomie der Universit\"{a}t Heidelberg, M\"{o}nchhofstr. 12-14, 69120 Heidelberg, \\ Germany \\
        }
\begin{document}

\date{Accepted 2014 August 26. Received 2014 August 26; in original form 2014 August 19}


\maketitle

\label{firstpage}

\begin{abstract}
We report the discovery of a circular mid-infrared shell around
the emission-line star Wray\,16-137 using archival data of the
{\it Spitzer} Space Telescope. Follow-up optical spectroscopy of
Wray\,16-137 with the Southern African Large Telescope revealed a
rich emission spectrum typical of the classical luminous blue
variables (LBVs) like P\,Cygni. Subsequent spectroscopic and
photometric observations showed drastic changes in the spectrum
and brightness during the last three years, meaning that
Wray\,16-137 currently undergoes an S\,Dor-like outburst. Namely,
we found that the star has brightened by $\approx$1 mag in the $V$
and $I_{\rm c}$ bands, while its spectrum became dominated by
Fe\,{\sc ii} lines. Taken together, our observations unambiguously
show that Wray\,16-137 is a new member of the family of Galactic
bona fide LBVs.
\end{abstract}


\section{Introduction}
\label{sec:intro}

\begin{figure}
\begin{center}
\includegraphics[width=8cm,angle=0]{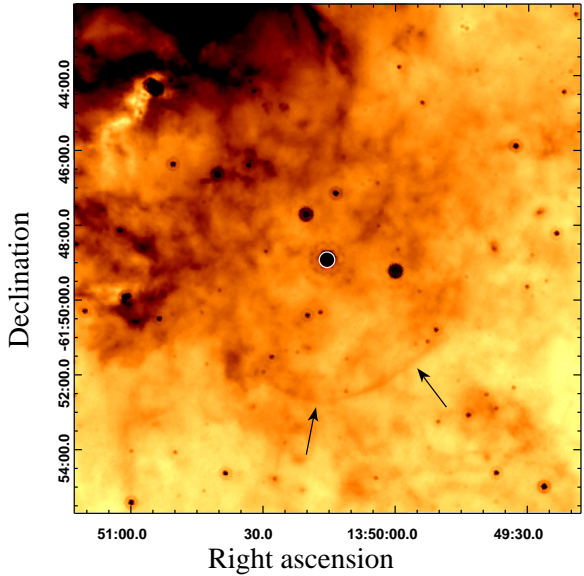}
\end{center}
\caption{ {\it Spitzer} MIPS $24 \, \mu$m image of a ring-like
nebula (indicated by arrows) around Wray\,16-137 (marked by a
white circle). A bright emission to the northeast of the nebula is
the \hii region GAL\,309.91+00.37.
    }
\label{fig:neb}
\end{figure}

During their life, some massive stars undergo a stage of drastic
changes in spectral appearance and brightness, which are
accompanied by episodes of enhanced mass loss (Humphreys \&
Davidson 1994). In this, so-called luminous blue variable (LBV;
Conti 1984), stage the stars change their visual brightness by one
or more magnitudes on time-scales of years, while their spectral
types vary between late O/early B and A/F supergiants. The nature
of the LBV-type activity remains unclear (see Vink 2012 for a
recent review), as well as whether the LBV stage is intermediate
between the main sequence and Wolf-Rayet stages or is an immediate
precursor of a supernova explosion (e.g. Langer et al. 1994;
Stothers \& Chin 1996; Groh, Meynet \& Ekstr\"{o}m 2013; Groh et
al. 2014; Smith \& Tombleson 2014). To some extent, this is
because the number of bona fide and candidate LBVs (cLBVs) has
remained quite sparse until recently (Clark, Larionov \& Arkharov
2005).

The number of known cLBVs in the Milky Way has greatly increased
with the advent of modern infrared telescopes (e.g. {\it Spitzer}
Space Telescope, {\it Wide-field Infrared Survey Explorer} [{\it
WISE}]), which resulted in the discovery of numerous compact
shells -- the distinctive characteristic of LBV and some other
evolved massive stars (Gvaramadze, Kniazev \& Fabrika 2010b;
Wachter et al. 2010). Follow-up spectroscopy of central stars of
these shells nearly doubled the number of Galactic cLBVs
(Gvaramadze et al. 2010a,b, 2012; Wachter et al. 2010, 2011;
Stringfellow et al. 2012a,b; Flagey et al. 2014). This burst of
discoveries conforms with the idea that LBVs could be the
descendants of not only the most massive and therefore very rare
stars (as it was generally believed), but also of the moderately
massive ($\sim$$20\,\msun$) and much more numerous ones (Smith et
al. 2011; Groh et al. 2013). However, none of the newly discovered
cLBVs have reported spectral and photometric variability strong
enough to call them bona fide LBVs (cf. Gvaramadze et al. 2010a,
2012). In this Letter, we present a first case of a new Galactic
bona fide LBV discovered through detection of a ring-like nebula
with {\it Spitzer} and follow-up spectroscopic and photometric
observations of its central star.

\section{Infrared nebula and its central star Wray\,16-137}
\label{sec:nebula}

The new nebula was discovered in the $24 \, \mu$m archival data of
the {\it Spitzer} Space Telescope obtained with the Multiband
Imaging Photometer for {\it Spitzer} (MIPS; Rieke et al. 2004)
within the framework of the 24 and 70 Micron Survey of the Inner
Galactic Disk with MIPS (Carey et al. 2009)\footnote{The discovery
was made serendipitously after our list of mid-IR nebulae detected
with {\it Spitzer} was already published (Gvaramadze et al.
2010b). One can expect that thorough inspection of complex
environments of star-forming regions -- where the majority of
massive stars are reside -- will disclose more new shells (e.g.
Gvaramadze et al. 2011).}. It appears as an incomplete ring-like
shell (see Fig.\,\ref{fig:neb}) of radius of $\approx$1.8 arcmin,
whose northeastern half is hidden by the bright emission
associated with the \hii region GAL\,309.91+00.37 (Caswell \&
Haynes 1987). The nebula can also be discerned in the
lower-resolution {\it WISE} 22\,$\mu$m image (Wright et al. 2010),
but it is invisible in the other three {\it WISE} bands (3.4, 4.6
and 12\,$\mu$m) and all (3.6, 4.5, 5.8 and $8.0\,\mu$m) images
obtained with the {\it Spitzer} Infrared Array Camera (IRAC; Fazio
et al. 2004) within the Galactic Legacy Infrared Mid-Plane Survey
Extraordinaire (Benjamin et al. 2003).

The nebula is centred on a point-like source known as
IRAS\,13467$-$6134 ($\alpha_{2000}$=$13^{\rm h} 50^{\rm m}
15\fs36$, $\delta_{J2000}$ =$-61\degr 48\arcmin 55\farcs2$). This
source is visible in all IRAC and {\it WISE} bands. Its optical
counterpart, named in the SIMBAD data base as Wray\,16-137 and
SS\,252, was identified as an emission-line star by Wray (1966)
and Stephenson \& Sanduleak (1977), respectively. In what follows,
we adopt the first of these two names. Wray\,16-137 was suspected
as an M supergiant with H$\alpha$ emission by MacConnell (1983).

The published optical photometry of Wray\,16-137 is very
uncertain. The SIMBAD data base quotes a visual magnitude of 15.5
without indication of the source of this value. The NOMAD
catalogue (Zacharias et al. 2004) gives $B$=14.1 mag and $V$=14.0
mag, based on the unpublished data, while the Yale/San Juan
Southern Proper Motion Catalog 4 (SPM4; Girard et al. 2011) lists
$B$=19.4 mag and $V$=16.0 mag. The discordance in the optical
photometry might be caused by the intrinsic variability of
Wray\,16-137 (see next section).

For the sake of completeness, we note that Wray\,16-137 is a
source of radio emission (Ricci et al. 2004).

\begin{figure*}
\begin{center}
\includegraphics[width=12cm,angle=270,clip=]{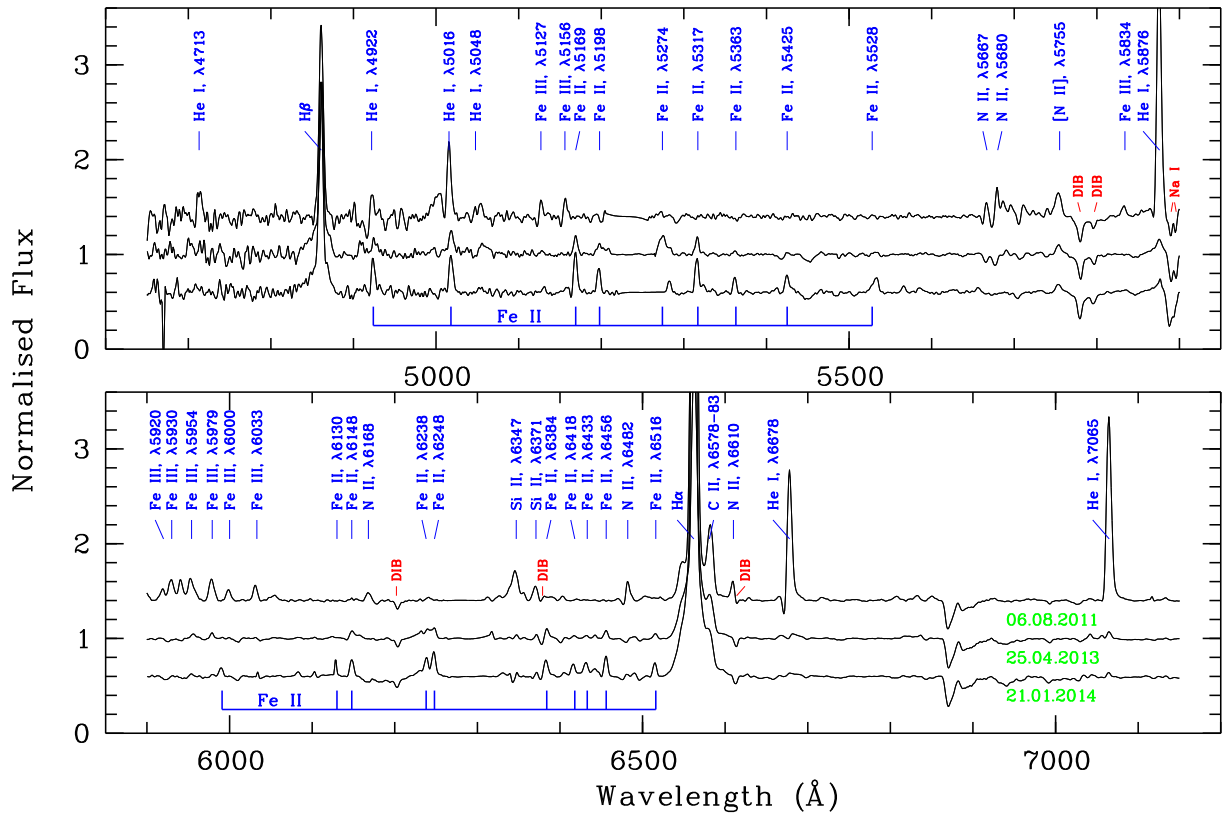}
\end{center}
\caption{Evolution of the (normalized) spectrum of Wray\,16-137
between 2011 August and 2014 January. The principal lines and most
prominent DIBs are indicated. For clarity, the spectra are offset
by 0.4 continuum flux unit.} \label{fig:spec}
\end{figure*}

\section{Wray\,16-137: a bona fide LBV}
\label{sec:LBV}

To clarify the nature of Wray\,16-137, we obtained its spectrum
with the Southern African Large Telescope (SALT; Buckley, Swart \&
Meiring 2006; O'Donoghue et al. 2006) on 2011 August 6, using the
Robert Stobie Spectrograph (RSS; Burgh et al. 2003; Kobulnicky et
al. 2003) in the long-slit mode with a $1.25\arcsec$ slit width.
The PG900 grating was used to cover the spectral range of
4200$-$7300 \AA \, with a final reciprocal dispersion of $0.97$
\AA \, pixel$^{-1}$ and full width at half-maximum (FWHM) spectral
resolution of 5.47$\pm$0.30 \AA. The RSS uses a mosaic of three
2048$\times$4096 CCDs and the final spatial scale for observations
was 0.253\arcsec\,pix$^{-1}$. Three 300\,sec spectra were taken.
The seeing during this and subsequent (see below) observations was
$\approx$2$\arcsec$$-$3$\arcsec$. A Xe lamp arc spectrum was taken
immediately after the science frames. A spectrophotometric
standard star was observed during twilight time for relative flux
calibration.

The primary reduction of the data was done with the SALT science
pipeline (Crawford et al. 2010). After that, the bias and gain
corrected and mosaiced long-slit data were reduced in the way
described in Kniazev et al. (2008).

The resulting normalized spectrum of Wray\,16-137 is presented in
Fig.\,\ref{fig:spec} (see the upper curve). It is dominated by
strong emission lines of H and He\,{\sc i} and numerous Fe\,{\sc
iii} emissions, some of which show P\,Cygni profiles. Further
emission lines in the spectrum are permitted singly ionized lines
of C, N and Si. No He\,{\sc ii} lines are present in the spectrum.
The only forbidden line detected is the line of [N\,{\sc ii}]
$\lambda 5755$. On the whole, the spectrum is very similar to
those of the bona fide LBVs P\,Cygni (Stahl et al. 1993) and
AG\,Car (near visual minimum; Groh et al. 2009) and several cLBVs
discovered with {\it Spitzer} and {\it WISE} through detection of
their associated mid-infrared shells (see figs 3-5 in Gvaramadze
et al. 2012). This spectrum along with the presence of the
circular shell around the star allow us to classify Wray\,16-137
as a cLBV.

It is likely that most, if not all, LBVs go through the long
(centuries or more) quiescent periods (Lamers 1986; Massey 2006),
during which they do not show major spectrophotometric
variability, and formally cannot be classified as genuine LBVs,
even though they possess LBV-like spectra and shells.
Nevertheless, if one is lucky enough, one can detect such
variability without having to wait for too long. Fortunately,
Wray\,16-137 provided us with this opportunity.

\begin{table}
  \caption{Photometry of Wray\,16-137.}
  \label{tab:phot}
  \renewcommand{\footnoterule}{}
  \begin{center}
  \begin{minipage}{\textwidth}
    \begin{tabular}{lccc}
      \hline
      Date & $B$ & $V$ & $I_{\rm c}$ \\
      \hline
      2007$^{(1)}$ & -- & 16.02$\pm$0.04 & -- \\
      2011 August 6$^{(2)}$ & -- & 15.24$\pm$0.03 & -- \\
      2012 May 6$^{(3)}$ & 17.98$\pm$0.10 & 15.05$\pm$0.04 & 10.06$\pm$0.01 \\
      2013 January 14$^{(3)}$ & 18.26$\pm$0.10 & 14.88$\pm$0.04 & 9.77$\pm$0.01 \\
      2013 April 25$^{(2)}$ & -- & 14.79$\pm$0.03 & -- \\
      2014 January 19$^{(3)}$ & 17.75$\pm$0.10 & 14.36$\pm$0.03 & 9.26$\pm$0.01 \\
      2014 January 21$^{(2)}$ & -- & 14.40$\pm$0.03  & -- \\
      2014 April 23$^{(3)}$ & 17.39$\pm$0.15 & 14.18$\pm$0.03 & 9.04$\pm$0.03 \\
      \hline
    \end{tabular}
    \end{minipage}
    \end{center}
     (1) SPM4; (2) SALT; (3) 76-cm telescope.
    \end{table}

To search for spectral variability of Wray\,16-137, we obtained
two additional spectra with the SALT using the same spectral
setup. These spectra, taken on 2013 April 25 and 2014 January 21,
are shown in Fig.\,\ref{fig:spec} along with the first spectrum.
The first look on them reveals that by 2014 January the He\,{\sc
i} emission lines had almost disappeared, which indicates that the
stellar effective temperature decreased during the last three
years. (A detailed spectral analysis of Wray\,16-137 is currently
underway and will be presented elsewhere.) This conclusion is
reinforced by the disappearance of the Fe\,{\sc iii} emission
lines, major changes in other temperature sensitive lines like C,
N and Si, and the appearance of numerous singly ionized iron
emission lines (some of which show P\,Cygni profiles).

The [N\,{\sc ii}] $\lambda 5755$ line also has changed
significantly. The FWHM and heliocentric radial velocity of this
line could be used as a measure of the stellar wind (Crowther,
Hillier \& Smith 1995) and systemic (Stahl et al. 2001)
velocities, $v_\infty$ and  $v_{\rm sys}$, respectively. After
correction for instrumental width, the FWHMs of 8.92$\pm$0.18 \AA
\, (2011), 7.46$\pm$0.31 \AA \, (2013) and 7.73$\pm$0.30 \AA \,
(2014) correspond to $v_\infty$ of 367$\pm$18, 264$\pm$22 and
285$\pm$22 $\kms$, respectively. As expected, $v_\infty$ decreased
as the star became cooler. For $v_{\rm sys}$ we derived a mean
value of $-$33$\pm$3 $\kms$ based on all three spectra.

To detect photometric variability of Wray\,16-137, we determined
its $B, V$ and $I_{\rm c}$ magnitudes on CCD frames obtained with
the the 76-cm telescope of the South African Astronomical
Observatory in 2012$-$2014. We used an SBIG ST-10XME CCD camera
equipped with $BVI_{\rm c}$ filters of the Kron-Cousins system
(see e.g. Berdnikov et al. 2012). Absolute flux calibration is not
feasible with SALT because the unfilled entrance pupil of the
telescope moves during the observations. However, we were able to
calibrate our spectra and synthesize their $V$ magnitudes (cf.
Kniazev et al. 2005) using a foreground star on the slit as a
secondary standard, whose photometry was determined from our CCD
frames. The results are presented in Table\,\ref{tab:phot}. To
this table we also added the $V$ magnitude (measured on CCD frames
in 2007) from the SPM4 catalogue (Girard et al. 2011)\footnote{The
$B$ magnitude of Wray\,16-137 in this catalogue was extrapolated
from the $JHK_{\rm s}$ photometry of the Two-Micron All Sky Survey
and therefore is less reliable.}. One can see that Wray\,16-137
monotonically brightened in the $V$ and $I_{\rm c}$ bands during
the last three years with the net increase of $\approx$1 mag.
Changes in $B$ are less secure because the weakness of the star in
this band causes larger errors of measurements. The brightening in
the $V$ band would be even more spectacular ($\approx$2 mag!) if
one takes into account the SPM4 photometry.

Taken together, our observations unambiguously show that
Wray\,16-137 is a bona fide LBV, which now experiences an
S\,Dor-type outburst and is on the way to visual maximum.

\section{Discussion and conclusion}

\begin{figure}
\begin{center}
\includegraphics[width=8cm,angle=0]{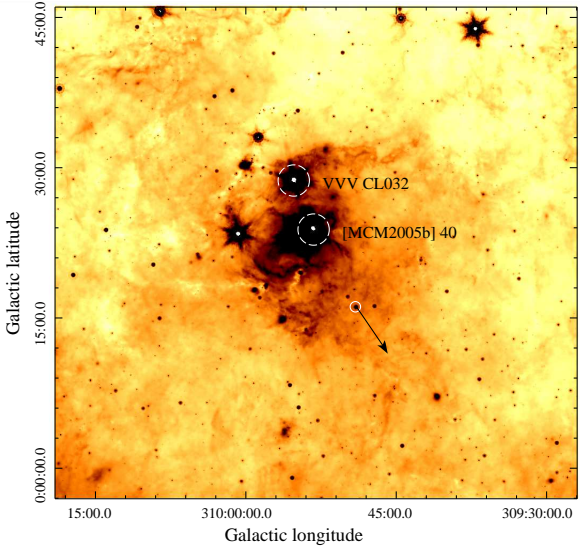}
\end{center}
\caption{ {\it Spitzer} MIPS $24 \, \mu$m image of the \hii region
GAL\,309.91+00.37 and two highly reddened clusters (marked by
dashed circles). The arrow shows the direction of motion of
Wray\,16-137 (marked by a white circle), as suggested by the
proper motion measurement for this star. See text for details. At
the distance of 4 kpc, 1 arcmin corresponds to $\approx$1.15 pc.}
\label{fig:hii}
\end{figure}

To estimate the reddening towards Wray\,16-137 and thereby to
constrain its absolute visual magnitude, $M_V$, we matched the
dereddened spectral slope of this star with those of stars of
similar effective temperature, $T_{\rm eff}$. Using the Stellar
Spectral Flux Library by Pickles (1998) and assuming that $T_{\rm
eff}$ of Wray\,16-137 in the hot phase (i.e. in 2011 August) is
similar to that of P\,Cygni ($\approx$18\,000 K; Najarro, Hillier
\& Stahl 1997), we found a colour excess of
$E(B$$-$$V)$$\approx$3.7 mag (this estimate only slightly depends
on the assumed $T_{\rm eff}$; see Gvaramadze et al. 2012). With
this $E(B$$-$$V)$ and the standard ratio of total to selective
extinction $R_V$=3.1, we obtained $A_V$$\approx$11.5 mag and
DM$+$$M_V$$\approx$3.8 mag, where $A_V$ is the $V$-band extinction
and DM is the distance modulus.

To constrain DM, we note that the line of sight towards
Wray\,16-137 first crosses the Carina-Sagittarius arm at $\sim$2
kpc and then tangentially crosses the Crux-Scutum one at
$\approx$4$-$10 kpc. Placing Wray\,16-137 in the first of these
two arms (DM$\approx$11.5 mag) would imply $M_V$$\approx$$-$7.7
mag, while in the next arm out (DM$\approx$13$-$15 mag) $M_V$
would be between $\approx$$-$9 and $-$11 mag. Correspondingly, the
luminosity of Wray\,16-137 would be $\log(L/\lsun)$$\approx$5.6
and 6.2$-$7.0; here we assumed that the bolometric correction of
Wray\,16-137 is equal to that of P\,Cygni, $-$1.54 mag (Najarro,
personal communication). In the first case, the luminosity of
Wray\,16-137 would be comparable to that of P\,Cigni of 5.7
(Najarro et al. 1997), while in the second one Wray\,16-137 would
be one of the most luminous Galactic LBVs.

The high $A_V$ towards Wray\,16-137 is suggestive of the longer
distance. The longer distance should also be accepted if
Wray\,16-137 is associated with the \hii region GAL\,309.91+00.37
(see Figs\,\ref{fig:neb} and \ref{fig:hii}), which is located in
the Crux-Scutum arm (Caswell \& Haynes 1987).
Table\,\ref{tab:prop} lists the components of the peculiar
transverse velocity (in Galactic coordinates), $v_{\rm l}$ and
$v_{\rm b}$, the peculiar radial velocity, $v_{\rm r}$, and the
total space velocity, $v_\ast$, of Wray\,16-137\footnote{To derive
these velocities, we used the Galactic constants $R_0 = 8.0$ kpc
and $\Theta _0 =240 \, \kms$ (Reid et al. 2009) and the solar
peculiar motion $(U_{\odot},V_{\odot},W_{\odot})=(11.1,12.2,7.3)
\, \kms$ (Sch\"onrich, Binney \& Dehnen 2010). For the error
calculation, only the errors of the proper motion and the systemic
velocity measurements were considered.}, derived from the proper
motion measurement given in the SPM4 catalogue (Girard et al.
2011): $\mu _\alpha \cos \delta$=$-$10.08$\pm$4.89 mas\,${\rm
yr}^{-1}$, $\mu _\delta$=$-$11.22$\pm$5.28 mas\,${\rm yr}^{-1}$.
For the sake of illustration, we adopted two distances: 2 and 4
kpc. Taken at face value, the obtained velocities imply that
Wray\,16-137 is a runaway star moving away from the geometric
centre of the \hii region. Interestingly, the SIMBAD data base
indicates two highly-reddened clusters, [MCM2005b]\,40 (Mercer et
al. 2005) and VVV\,CL032 (Borissova et al. 2011), within the
boundaries of GAL309.91+00.37. Although the distances to these
clusters and their stellar contents are unknown, it is likely that
they power the \hii region and that one of them might be the
parent cluster of Wray\,16-137. Upcoming high-precision proper
motion and parallax measurements with the space astrometry mission
Gaia would allow to derive the distance to Wray\,16-137 and to
prove whether it could be associated with GAL\,309.91+00.37.

\begin{table}
\caption{Peculiar transverse (in Galactic coordinates) and radial
velocities, and the total space velocity of Wray\,16-137 for two
adopted distances (see text for details).} \label{tab:prop}
\centering
\begin{tabular}{ccccccc}
\hline
$d$ & $v_{\rm l}$ & $v_{\rm b}$ & $v_{\rm r}$ & $v_\ast$ \\
(kpc) & ($\kms$) & ($\kms$) & ($\kms$) & ($\kms$) \\
\hline
2 & $-$46$\pm$47 & $-$75$\pm$50 & $-$5$\pm$3 & 88$\pm$49 \\
4 & $-$107$\pm$93 & $-$157$\pm$100 & 17$\pm$3 & 191$\pm$97 \\
\end{tabular}
\end{table}

To conclude, further spectroscopic and photometric observations of
this interesting star with better cadence and higher spectral
resolution and coverage are very desirable.

\section{Acknowledgements}
We are grateful to I.D.~Howarth (the referee) for useful
suggestions on the manuscript. Some observations reported in this
paper were obtained with the Southern African Large Telescope
(SALT), programmes \mbox{2010-1-RSA\_OTH-001},
\mbox{2013-1-RSA\_OTH-014} and \mbox{2013-2-RSA\_OTH-003}. AYK
acknowledges support from the National Research Foundation (NRF)
of South Africa. This work is based in part on archival data
obtained with the {\it Spitzer} Space Telescope, which is operated
by the Jet Propulsion Laboratory, California Institute of
Technology under a contract with NASA, and has made use of the
NASA/IPAC Infrared Science Archive, which is operated by the Jet
Propulsion Laboratory, California Institute of Technology, under
contract with the National Aeronautics and Space Administration,
the SIMBAD database and the VizieR catalogue access tool, both
operated at CDS, Strasbourg, France.


\begin{thebibliography}{}
%
\bibitem{} Benjamin R. A. et al., 2003, PASP, 115, 953
\bibitem{} Berdnikov L. et al., 2012, Astronomy Reports, 56, 290
\bibitem{} Borissova J. et al., 2011, A\&A, 532, A131
\bibitem{} Buckley D. A. H., Swart G. P., Meiring J. G., 2006, SPIE, 6267, 32
\bibitem{} Burgh E. B., Nordsieck K. H., Kobulnicky H. A., Williams T. B., O'Donoghue D., Smith M. P., Percival J. W., 2003, SPIE, 4841, 1463
\bibitem{} Carey S. J. et al., 2009, PASP, 121, 76
\bibitem{} Caswell J. L., Haynes R. F., 1987, A\&A, 171, 261
\bibitem{} Clark J. S., Larionov V. M., Arkharov A., 2005, A\&A, 435, 239
\bibitem{} Conti P. S., 1984, in Maeder A., Renzini A., eds, Observational Tests of the Stellar Evolution Theory. Reidel, Dordrecht, p. 233
\bibitem{} Crawford S. M. et al., 2010, SPIE, 7737
\bibitem{} Crowther P. A., Hillier D. J., Smith L. J., 1995, A\&A, 293, 172
\bibitem{} Fazio G. G. et al., 2004, ApJS, 154, 10
\bibitem{} Flagey N., Noriega-Crespo A., Petric A. O., Geballe T. R., 2014, AJ, 148, 34
\bibitem{} Girard T. M. et al., 2011, AJ, 142, 15
\bibitem{} Groh J. H., Hillier D. J., Damineli A., Whitelock P. A., Marang F., Rossi C., 2009, ApJ, 698, 1698
\bibitem{} Groh J. H., Meynet G., Ekstr\"{o}m S., 2013, A\&A, 550, L7
\bibitem{} Groh J., Meynet G., Ekstr\"{o}m S., Georgy C., 2014, A\&A, 564, A30
\bibitem{} Gvaramadze V. V., Kniazev A. Y., Fabrika S., Sholukhova O., Berdnikov L. N., Cherepashchuk A. M., Zharova
A. V., 2010a, MNRAS, 405, 520
\bibitem{} Gvaramadze V. V., Kniazev A. Y., Fabrika S., 2010b, MNRAS, 405, 1047
\bibitem{} Gvaramadze V. V., Kniazev A. Y., Kroupa P., Oh S., 2011, A\&A, 535, A29
\bibitem{} Gvaramadze V. V. et al., 2012, MNRAS, 421, 3325
\bibitem{} Humphreys R. M., Davidson K., 1994, PASP, 106, 1025
\bibitem{} Kniazev A. Y. et al., 2005, AJ, 130, 1558
\bibitem{} Kniazev A. Y. et al., 2008, MNRAS, 388, 1667
\bibitem{} Kobulnicky H. A., Nordsieck K. H., Burgh E. B., Smith M. P., Percival J. W., Williams T. B., O'Donoghue D., 2003, SPIE, 4841, 1634
\bibitem{} Lamers H. J. G. L. M., 1986, in  de Loore C. W. N., Willis A. J., Laskarides P., eds, Luminous Stars and Associations in Galaxies. Reidel, Dordrecht, p. 157
\bibitem{} Langer N., Hamann W.-R., Lennon M., Najarro F., Pauldrach A. W. A., Puls J., 1994, A\&A, 290, 819
\bibitem{} MacConnell D. J., 1983, Rev. Mex. Astron. Astrofis., 8, 39
\bibitem{} Massey P., 2006, ApJ, 638, L93
\bibitem{} Mercer E. P. et al., 2005, ApJ, 635, 560
\bibitem{} Najarro F., Hillier D. J., Stahl O., 1997, A\&A, 326, 1117
\bibitem{} O'Donoghue D. et al., 2006, MNRAS, 372, 151
\bibitem{} Pickles A. J., 1998, PASP, 110, 863
\bibitem{} Reid M. J., Menten K. M., Zheng X. W., Brunthaler A., Xu Y., 2009, ApJ, 705, 1548
\bibitem{} Ricci R. et al., 2004, MNRAS, 354, 305
\bibitem{} Rieke G. H. et al., 2004, ApJS, 154, 25
\bibitem{} Sch\"onrich R., Binney J., Dehnen W., 2010, MNRAS, 403, 1829
\bibitem{} Smith N., Li W., Silverman J. M., Ganeshalingam M., Filippenko A. V., 2011, MNRAS, 415, 773
\bibitem{} Smith N., Tombleson R., 2014, preprint (arXiv:1406.7431)
\bibitem{} Stahl O., Mandel H., Wolf B., Gaeng Th., Kaufer A., Kneer R., Szeifert Th., Zhao F., 1993, A\&AS, 99, 167
\bibitem{} Stahl O. et al., 2001, A\&A, 375, 54
\bibitem{} Stephenson C. B., Sanduleak N., 1977, ApJS, 33, 459
\bibitem{} Stothers R. B., Chin C.-W., 1996, ApJ, 468, 842
\bibitem{} Stringfellow G. S., Gvaramadze V. V., Beletsky Y., Kniazev A. Y., 2012a, in Richards M. T., Hubeny I., eds, Proc. IAU Symp. 282, From
Interacting Binaries to Exoplanets: Essential Modeling Tools.
Cambridge Univ. Press, Cambridge, p. 267
\bibitem{}Stringfellow G. S., Gvaramadze V. V., Beletsky Y., Kniazev A. Y., 2012b, in Drissen L., St-Louis N., Robert C., Moffat A. F. J.,
eds, ASP Conf. Ser. Vol. 465, Four Decades of Massive Star
Research -- A Scientific Meeting in Honor of Anthony J. Moffat.
Astron. Soc. Pac., San Francisco, p. 514
\bibitem{} Vink J. S., 2012, in Davidson K., Humphreys R. M., eds, Astrophys. \& Sp. Sci. Library Vol. 384, Eta Carinae and the
Supernova Impostors, New York, p. 221
\bibitem{} Wachter S., Mauerhan J. C., van Dyk S. D., Hoard D. W., Kafka S., Morris P. W., 2010, AJ, 139, 2330
\bibitem{} Wachter S., Mauerhan J., van Dyk S., Hoard D. W., Morris P., 2011, Bull. Soc. R. Sci. Li\`{e}ge, 80, 322
\bibitem{} Wray J. D., 1966, PhD thesis, Northwestern Univ.
\bibitem{} Wright E. L. et al., 2010, AJ, 140, 1868
\bibitem{} Zacharias N., Monet D. G., Levine S. E., Urban S. E., Gaume R., Wycoff G. L., 2004, BAAS, 36, 1418
%
\end{thebibliography}
\end{document}